\documentclass[A4paper,useAMS,usenatbib]{mn2e}                                                               
\usepackage{graphicx,amssymb,color}                                                                          
\usepackage{lastpage}                                                                                        
\usepackage[fleqn]{amsmath}                                                                                  
\usepackage[normalem]{ulem}                                                                                  
\input psfig

\title[Exoplanet frequency with PLATO]
{Prospects for detecting decreasing exoplanet frequency with main sequence age using PLATO}
\author[Veras, Brown, Mustill, \& Pollacco]{Dimitri Veras$^{1}$\thanks{E-mail: d.veras@warwick.ac.uk}, 
David J.A. Brown$^{1,2}$, Alexander J. Mustill$^{3}$, Don Pollacco$^{1}$\\
$^{1}$Department of Physics, University of Warwick, Coventry CV4 7AL, United Kingdom\\
$^{2}$Astrophysics Research Centre, School of Mathematics \& Physics, Queen's University Belfast, University Road, Belfast, BT7 1NN\\
$^{3}$Lund Observatory, Department of Astronomy and Theoretical Physics, Lund University, Box 43, SE-221 00, Lund, Sweden
}

\begin{document}

%\date{Accepted. Received; in original form }

\pagerange{\pageref{firstpage}--\pageref{lastpage}} \pubyear{2015} 

\maketitle

\label{firstpage}

\begin{abstract}
The space mission {\it PLATO} will usher in a new era of exoplanetary science by expanding our current inventory of transiting systems and constraining host star ages, which are currently highly uncertain. This capability might allow {\it PLATO} to detect changes in planetary system architecture with time, particularly because planetary scattering due to Lagrange instability may be triggered long after the system was formed.  Here, we utilize previously published instability timescale prescriptions to determine {\it PLATO}'s capability to detect a trend of decreasing planet frequency with age for systems with equal-mass planets.  For two-planet systems, our results demonstrate that {\it PLATO} may detect a trend for planet masses which are at least as massive as super-Earths. For systems with three or more planets, we link their initial compactness to potentially detectable frequency trends in order to aid future investigations when these populations will be better characterized.
\end{abstract}

\begin{keywords}
planets and satellites: dynamical evolution and stability -- techniques: photometric -- stars: evolution --  stars: solar-type -- celestial mechanics -- telescopes
\end{keywords}

\section{Introduction}

Nearly all exoplanets orbit stars whose ages are poorly constrained.  This 
situation is unfortunate
because the potential to learn about the dynamical evolution of planetary systems 
through the host star's evolution has yet to be realized.  Accurate stellar
ages may provide key constraints and insights on the formation and fate of planets.

Tidally-influenced, hot exoplanets highlight the importance of this link.
Hundreds of confirmed planets have orbital periods less than 10 days,
where tidal interactions between the planet and star can affect the
planet's orbital parameters in a myriad of ways \citep[e.g. see Fig. 4 of][]{ogilvie2014}.
Because the timescales for shrinking an orbit are sensitively dependent
on the orbital period, accurate stellar ages would crucially constrain the dynamical
histories of these systems. Further, the details of tidal theory are controversial and 
recent analyses \citep[e.g.][]{efrmak2013} demonstrate the dangers of adopting
the heretofore widely-used constant quality factor.  Adding stellar age
to the list of known system parameters would enable greater resolution on this topic.

However, the insights obtained from accurate stellar ages are not restricted to 
substellar companions with such close orbits.  Single planets or brown dwarfs 
which are out of the reach of stellar tides (with semimajor axes beyond about 0.1 au)
but within the ellipsoid of the gravitational influence of the parent star 
within the Milky Way (within about $10^5$ au) could attain currently-observed orbital 
properties from (i) a relic planet formation process such as core accretion, gravitational
instability or disc migration \citep[see][for recent reviews]{baretal2013,heletal2013}, 
(ii) a relic star formation process such as star-planet scattering in birth clusters
\citep{phamut2006,spuetal2009,maletal2011,parqua2012,haoetal2013},
(iii) gravitational interaction with other planets during a star's main sequence
lifetime \citep[see][for a recent review]{davetal2014} 
or/and (iv) perturbations from Galactic field star flybys 
\citep{zamtre2004,freetal2006,boletal2012,vermoe2012}
and Galactic tides \citep{heitre1986,brasser2001,vereva2013}.

All these processes are time-dependent and hence could benefit 
from accurate stellar ages.  Techniques are being developed to 
constrain stellar ages \citep{lebgou2014} to within 10 per cent
of a typical main sequence lifetime of a Solar-mass star by
utilizing astroseismology.  Although these techniques can
be applied to high-precision photometric data taken from the
{\it CoRoT} \citep{bagetal2002} and {\it Kepler} \citep{boretal2010,kocetal2010}
space missions, the number of suitable stars from these missions is
just a handful.

The {\it PLATO} space mission \citep{rauetal2014}, due for launch in 2024, 
will achieve similarly tight age constraints, but for a much larger
sample of stars.  One of the main aims of {\it PLATO} is to achieve 
10\% precision in the age of 20,000 main sequence stars of spectral 
types F5-K7.  In order to achieve this goal, {\it PLATO} will produce internal
stellar mass distributions for different stars by combining 
radius measurements derived from {\it Gaia} data with the oscillation 
frequencies observed in the {\it PLATO} light curves.

Other techniques which will be used to derive stellar ages, 
such as gyrochronology, 
will be assessed and calibrated from this ``primary" sample and then 
applied to the hundreds of thousands of main sequence stars that 
{\it PLATO} will observe during the course of its 6 year mission 
(albeit at a lower age precision). Moreover, {\it PLATO} will, for the 
primary sample, allow for direct testing of stellar evolution tracks.

Such a large sample will enable us to detect trends in exoplanetary
science that remain currently hidden in the noise.  One of these trends
is frequency of planets versus main sequence age.  Not all systems 
with multiple planets will remain stable over their main sequence 
lifetimes.  In our own Solar system, Mercury has a one to few 
per cent chance of destabilizing the inner planets \citep{lasgas2009,zeebe2015}.
Other systems are closer to instability; Section 6.3 of \cite{veretal2013}
discuss specific examples for planets discovered by Doppler radial velocity
with semimajor axes that exceed 1 au.  Overall, the frequency of planets 
might decrease with time.

The trend for planets detected by transit photometry within 1 au but beyond the 
limit at which tidal forces cease to become significant (at about 0.1 au) should 
not be qualitatively different.  On 9 Jul 2015,
the Exoplanets Data Explorer (http://exoplanets.org) listed 509 substellar objects 
in this semimajor axis range that are components of multiple-planet systems.
Overall, \cite{fanmar2012} and \cite{baljoh2014} both estimate that about half
of {\it Kepler} systems are multiple-planet systems, the latter considering
only M dwarfs.  Seven examples of potential planets in multi-planet systems
are the seven candidate planets of KIC 11442793, all of which transit
between 0.074 au and 1.0 au but do not have measured masses \citep{cabetal2014}.
In fact, those authors conclude that the two outermost candidates are 
indeed planets due to dynamical stability constraints on their masses, 
highlighting how close to instability some of these systems might reside.

In this brief paper, we combine stability prescriptions from 
previous investigations with the detection capabilities of 
the {\it PLATO} mission.  Section 2 identifies these prescriptions
and Section 3 performs the analysis.  We conclude in Section 4.

\section{Stability of planetary systems}

Two-planet systems represent a robust starting point with which to study decreasing
planet frequency with time along the main sequence.  
Observers have identified hundreds of these systems, and theorists have established
analytical constraints which would otherwise be unattainable with additional planets.

\subsection{Two-planet stability}

A two-planet system is a generally-unsolvable three-body problem.  However, 
\cite{marsaa1975} and \cite{marboz1982} identified allowable regions of motion in this 
problem based on the angular momentum and energy of the system.  This 
categorization proved particularly valuable after the confident discovery of 
extrasolar planets 
\citep{wolfra1992} for assessing their stability.  Many subsequent studies,
summarized in \cite{georgakarakos2008}, considered relevant special cases and extensions.

The allowable regions of motions are partially partitioned by the concept of Hill stability.
Two planets are said to be {\it Hill stable} if their orbits never cross.  This requirement
does not preclude the inner planet from crashing into the star, nor the outer planet
from escaping from the system.  If the planets do remain ordered and do not suffer a
collision nor ejection, then the system is said to be {\it Lagrange stable}.
In Hill unstable cases, when the initial orbits do cross, then
the outcome is uncertain: the system could destabilize or alternatively 
the planets might remain inside of a mean motion resonance (like Neptune and Pluto).
This latter possibility presupposes the planets were originally captured into a resonance,
most likely due to protoplanetary disc dissipation, but not necessarily \citep{rayetal2008}.

Nevertheless, in most cases (as a fraction of available parameter space), planets which 
cross orbits will become violently unstable.  
In this respect, the critical separation 
$\Delta_{\rm H}$\footnote{The boundary of the Hill radius of a single planet is a related 
concept \citep[see Appendix B of][]{peawya2014}, but not equivalent to $\Delta_{\rm H}$.} 
of the initial planetary orbits that determines
the Hill stability boundary has occasionally been treated, erroneously, as a type of 
global stability boundary. \cite{marzari2014} has demonstrated why Hill stability
should not be treated as a global boundary through a frequency map analysis.
In some cases though, the Hill stability boundary may be considered
a dividing line between ``quick instability''
and ``slow instability''. This notion may be carried further: one can then define a
Lagrange stability boundary $\Delta_{\rm L} > \Delta_{\rm H}$ beyond which Lagrange
instability does not occur, or occurs on a timescale which exceeds the main sequence stellar lifetime.

The lack of quantification in these statements results from a dearth of dedicated 
studies of Lagrange instability.
We summarize previous investigations as follows:
Chapter 11 of \citeauthor{marchal1990} (\citeyear{marchal1990}), \cite{anosova1996} and \cite{lietal2009} 
focused on the escape of the outer body; \cite{khokuz2011} considered the mass 
dependence; \cite{decetal2013} linked Lagrange instability (albeit defined slightly 
differently) with mean motion resonance overlap; and \cite{bargre2006,bargre2007}, 
\cite{rayetal2009}, \cite{kopbar2010},
\cite{veretal2013} and \cite{vermus2013} 
evaluated the extent of Lagrange instability beyond the Hill stability limit.
Recently \cite{petrovich2015} unveiled a fairly general Lagrange instability criterion
for two unequal-mass, highly eccentric and nearly coplanar planets that performs
significantly better than previous widely-used criteria from stellar dynamics
\citep{eggkis1995,maraar2001}.
Although excellent analytic approximations to $\Delta_{\rm H}$ exist \citep{donnison2011}, 
the same is not true for $\Delta_{\rm L}$.  One must find $\Delta_{\rm L}$ empirically through
simulations.  However, simulations have revealed that this Lagrange stability boundary
is nebulous.

\cite{vermus2013} found that although $\Delta_{\rm L}$ is difficult to constrain,
the minimum time to Lagrange instability is well-established empirically as
a function of planet mass only, for equal-mass planets.  They defined the Lagrange
instability timescale as the time at which either the inner planet collided with the
stellar surface or when the outer planet's orbit became hyperbolic.  If $x_{\rm uns}$ 
represents the minimum number of initial orbits of the inner planet before the onset of 
Lagrange instability, then

\begin{equation}
\log_{10}{\left(x_{\rm uns}\right)} \sim \left(5.23 \pm 0.04\right) \left(\frac{\mu}{M_{\rm J}/M_{\odot}}\right)^{-0.181\pm 0.003}
,
\label{uns}
\end{equation}

\noindent{}where $\mu$ is the planet/star mass ratio, for both planets, and $M_{\rm J}$ is
the mass of Jupiter.  This formula
appears to hold true for any separation in-between $\Delta_{\rm H}$ and $\Delta_{\rm L}$,
for, at least, orbits with initial eccentricities less than 0.3, and is valid for both 
planets and brown dwarfs.  The equation is based on a fit down to masses of $0.1 M_{\rm J}$,
and will be used directly in conjunction with {\it PLATO} sensitivities in Section 3.

\subsection{Stability of three or more planets}

In systems with three or more planets, there is no known analytical constraint on 
Hill stability, and the distinction between Hill instability and Lagrange instability
is rarely asserted.  Instead, researchers have relied on empirical fits to $N$-body
simulations to help characterize the potential future stability of a system.
Analytical motivation for establishing these fits is also rare. For an exception,
see Section 3 of \cite{zhoetal2007}, who demonstrate how an
analytical treatment from first principles can reproduce the same scalings 
obtained from their empirical fits.

The functional form predominately chosen for the global instability fit for
three or more planets is equivalent to \citep{chaetal1996,smilis2009,funetal2010,puwu2015}

\begin{equation}
\log_{10}\left({x_{\rm uns}}\right) \sim b \beta + c
,
\label{comlog}
\end{equation}

\noindent{}although slight variations exist \citep{fabqui2007,zhoetal2007,chaetal2008,quillen2011,musetal2014}.  In equation (\ref{comlog}), $b$ and $c$ are fitted constants and $\beta$ refers to the
number of \textit{mutual Hill radii} by which consecutive pairs of planets are separated.
A mutual Hill radius is a distance that is not defined uniformly throughout the literature 
(contrast, e.g., \citealt*{chaetal1996} and \citealt*{marwei2002} with \citealt*{smilis2009}, 
and see the discussion at
the end of Section 2 of \citealt*{musetal2014}).  Regardless, effectively, $\beta$ determines 
the extent of \textit{packing} in a planetary system 
\citep{barray2004,baretal2008,rayetal2009,vergae2015}, and can be expressed through masses
and semimajor axes.  Although the exact definition of $\beta$ varies \citep[see also the sentence
containing equation 5 in][]{puwu2015}, overall if $\beta$ is too small, then the system will
become unstable almost immediately.  Alternatively, if $\beta$ is too large, the system will
not experience instability during the star's main sequence lifetime.

Like Equation (\ref{uns}), Equation (\ref{comlog}) is applicable only for equal-mass planets.
However, Equation (\ref{comlog}) is more versatile, having been applied to systems with for
example 3, 5, 10 and even 20 planets.
The unequal-planet mass case greatly increases the available phase space to explore, and has so-far
largely been ignored (we do so here as well).  Although Equation (\ref{comlog}) may be more 
versatile, this general form does not appear to constrain stability as well as 
Equation (\ref{uns}), because the latter is for the specific case of two equal-mass planets.

Whereas $\beta$ typically represents a variable user-defined proxy for initial separations,
the values of $b$ and $c$ are considered to be fixed and are computed only after
a suite of $N$-body simulations has finished running.  Both the number of planets as well as the planet 
mass / star mass ratio under the assumption of initially circular orbits sets both $b$ and $c$; these 
values might also be weakly dependent on $a_1$ and on $\beta$.  In fact, the LHS of Equation
(\ref{comlog}) is more often expressed as an unscaled time.

The wide applicability of Equation (\ref{comlog}) for systems with at least three planets
is useful for our purposes.  Although that equation does not let us directly relate
\textit{PLATO} sensitivities with planetary mass, as in the two-planet case, the
relation provides an indirect pathway to do so depending on the architecture a reader
might consider.

\section{Link with PLATO}

Assume that {\it PLATO} will be able to constrain a star's age to $\Delta t_{\rm PLATO}$,
such that {\it PLATO}  will be able to distinguish $(t_{\rm MS}/\Delta t_{\rm PLATO})$ time bins 
over a star's main sequence lifetime $t_{\rm MS}$. At the level of approximation
we are aiming for, we need not require that the number of time bins is an integer,
and may also treat $\Delta t_{\rm PLATO}$ as constant in time.  With these assumptions, we can
then write the number of initial inner planetary orbits whose total duration encompasses 
one time bin as

\begin{equation}
x_{\rm min} \equiv 
\frac
{\sqrt{GM_{\star}}\Delta t_{\rm PLATO}}
{2\pi a_{1}^{3/2}}
.
\label{min}
\end{equation}

\noindent{}One can consider $x_{\rm min}$ to be the minimum number of orbits
needed to resolve a trend of planet frequency with stellar age.

However, no resolution is possible if the planets (i) fail to become unstable,
or (ii) all become unstable too early.  In the first case, instability
must occur before the start of the last time bin, such that we
require

\begin{equation}
x_{\rm uns} <
\left(
\frac{t_{\rm MS}}{\Delta t_{\rm PLATO}} - 1
\right)
x_{\rm min}
.
\label{max}
\end{equation}

\noindent{}This condition alone may be sufficient to detect a trend
of decreasing planet mass with stellar age.  If, however, the instability
occurs almost exclusively within the first time bin, then equation
(\ref{max}) is not restrictive enough.  In this case, we should also
impose the condition

\begin{equation}
x_{\rm uns} >
x_{\rm min}
.
\label{min}
\end{equation}

We choose to leave $\Delta t_{\rm PLATO}$ as an independent
variable to allow for the possibility of variations in the stellar constraints
on a star-by-star basis.  Further, we consider
9 different main sequence stellar masses ($0.7,0.8,0.9,1.0,1.1,1.2,1.3,1.4,1.5 M_{\odot}$),
corresponding to the range in which asteroseismology with PLATO can be effective at constraining 
stellar ages.  Because we are interested only in Lagrange instability on the main sequence
\footnote{Nearly all of the (transiting) planets assumed here will be engulfed by the star during
post-main-sequence evolution \citep{musvil2012,norspi2013,viletal2014}.
If the planets were further away from the star, then Lagrange instability could occur
during {\it either} the giant branch or white dwarf phases of stellar evolution
\citep{veretal2013}.}, 
we must compute the main sequence lifetime $t_{\rm MS}$
of these stars.  To do so, we use the {\sc sse} code \citep{huretal2000},
and assume Solar metallicity.
We obtain $t_{\rm MS} = 10948,7646,5607,4237,3355,$
and 2716 Myr for 
$M_{\star}=1.0,1.1,1.2,1.3,1.4,1.5M_{\odot}$, respectively.
The lower mass stars are on the main sequence for a duration greater
than the current age of the Universe.  Consequently, for added clarity,
we do not draw curves in the figures of this paper for the $0.8M_{\odot}$
and $0.9M_{\odot}$ tracks, as they are so similar to the $0.7M_{\odot}$
track.

\subsection{Two-planet systems}

First consider equation (\ref{max}) in the context
of two-planet systems.  Combining that with equation (\ref{uns})
yields the following specific condition on the planet/star
mass ratio as a function of $\Delta t_{\rm PLATO}$

\[
\frac{\mu}{M_J/M_{\odot}} \gtrsim 0.026 \bigg[1 + 0.10\log_{10}{}\bigg\lbrace
\]

\[
\ \ \ \ \ \ \ \ \ \ \ \ \
\left(
\frac{M_{\star}}{M_{\odot}}
\right)^{\frac{1}{2}}
\left(
\frac{a_1}{0.2 \ {\rm au}}
\right)^{-\frac{3}{2}}
\left(
\frac{\Delta t_{\rm PLATO}}{1 \ {\rm Gyr}}
\right)
\]

\[
\ \ \ \ \ \ \ \ \ \ \ \ \
\times
\left(
\frac{t_{\rm MS}}{\Delta t_{\rm PLATO}} - 1
\right)
\bigg\rbrace
\bigg]^{-5.55}
.
\]

\begin{equation}
\label{keyeq}
\end{equation}

%%%%%%%%%%%%%%%%% Figure eqmass
\begin{figure}
\centerline{
\psfig{figure=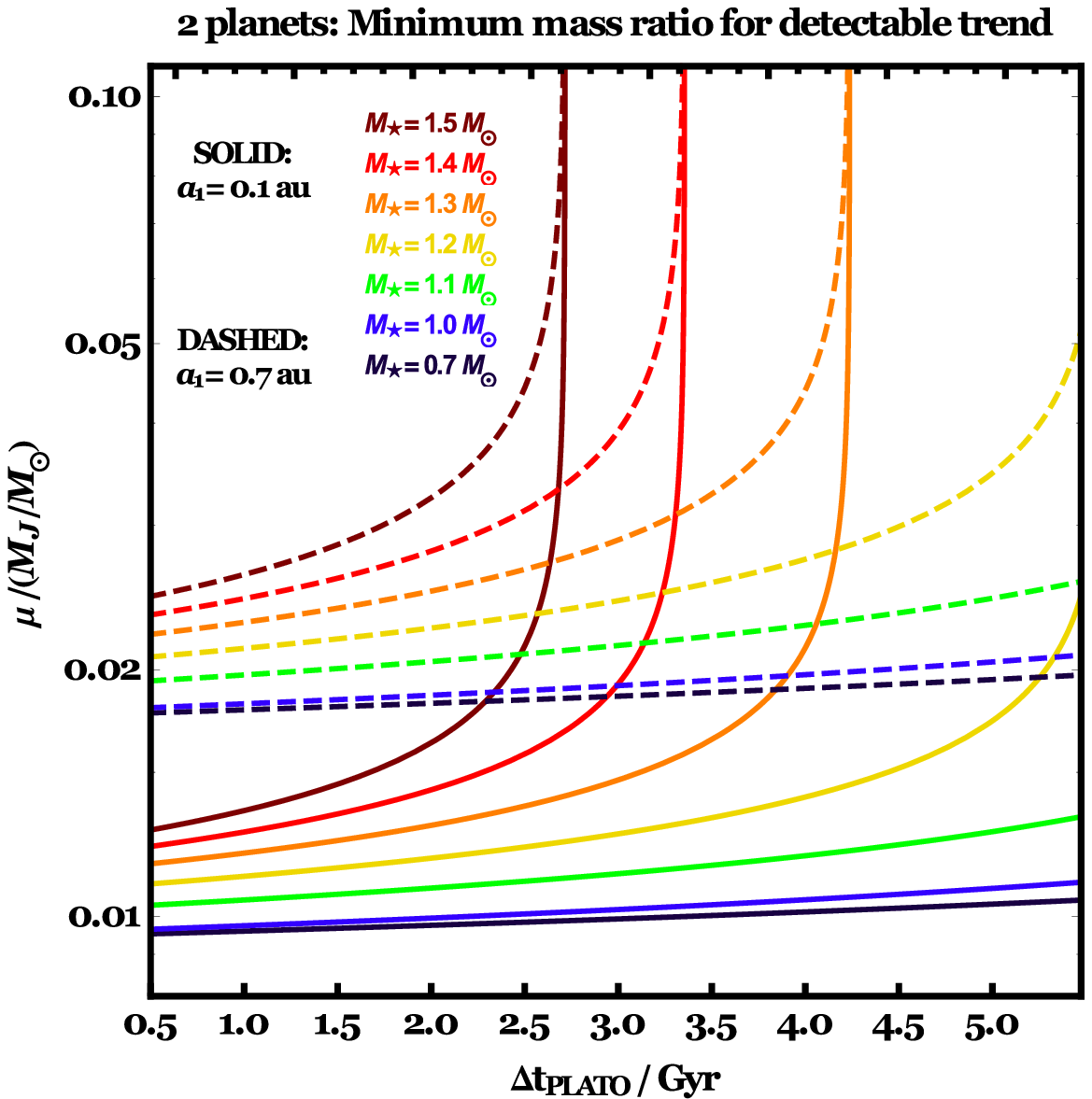,height=6.8cm,width=8.0cm} 
}
\centerline{
\psfig{figure=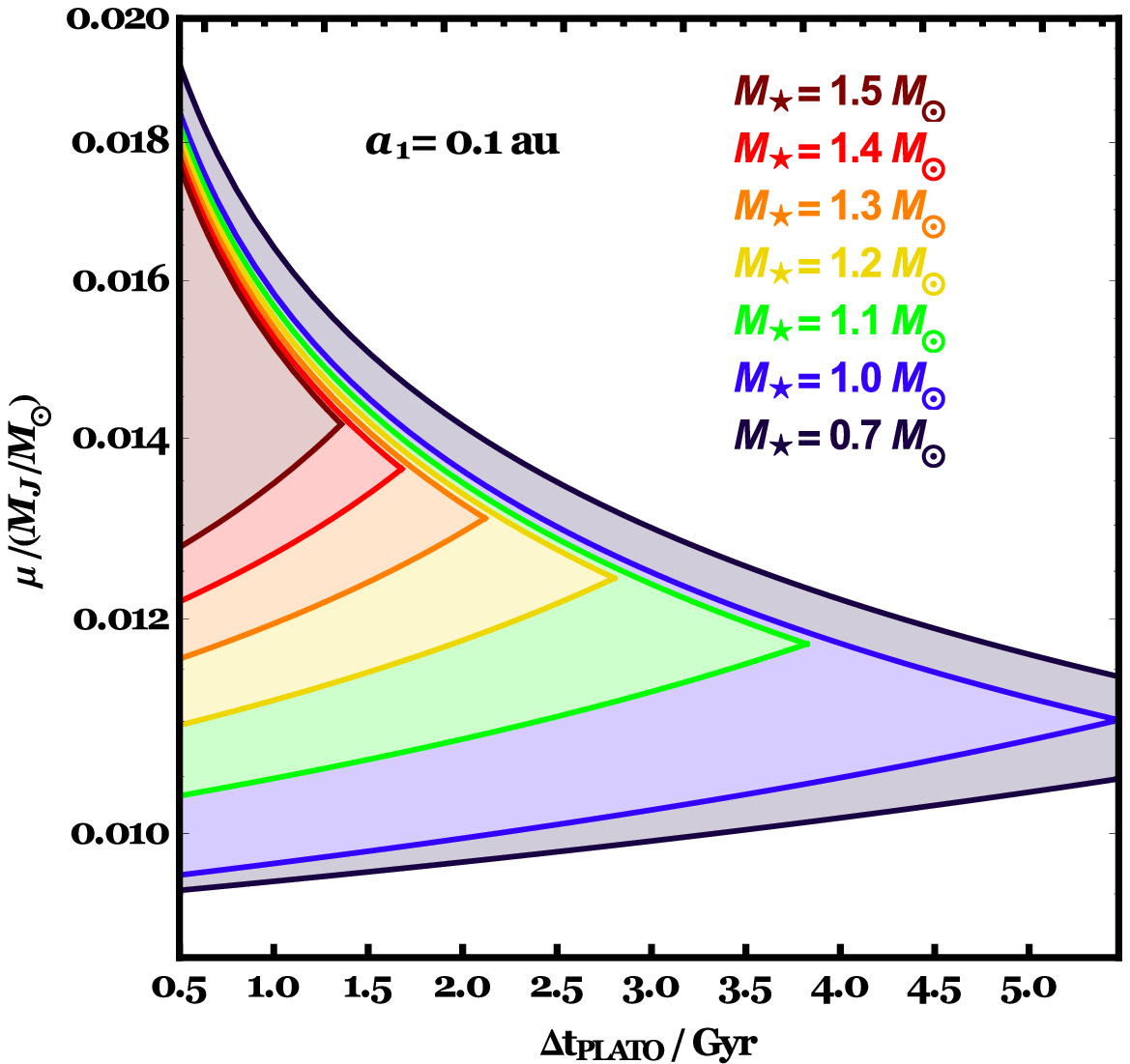,height=6.8cm,width=8.0cm} 
}
\caption{The minimum planet-star mass ratio $\mu$ for which {\it PLATO} can 
detect a decreasing trend of planet frequency versus time for packed, Hill-stable
two-planet systems.  
The $x$-axis refers to the (variable) magnitude of the stellar age
constraints {\it PLATO} may provide.  In the lower panel, the first time
bin is excluded, emphasizing its importance at detecting potential trends
(shaded regions only).
If ages are constrained to within 1 Gyr,
then a trend should be detectable for planets at least as massive as 
$10^{-2} M_{\rm J}$.
}
\label{Plan2}
\end{figure}
%%%%%%%%%%%%%%%%% Figure eqmass

%%%%%%%%%%%%%%%%% Figure eqmass
\begin{figure}
\centerline{
\psfig{figure=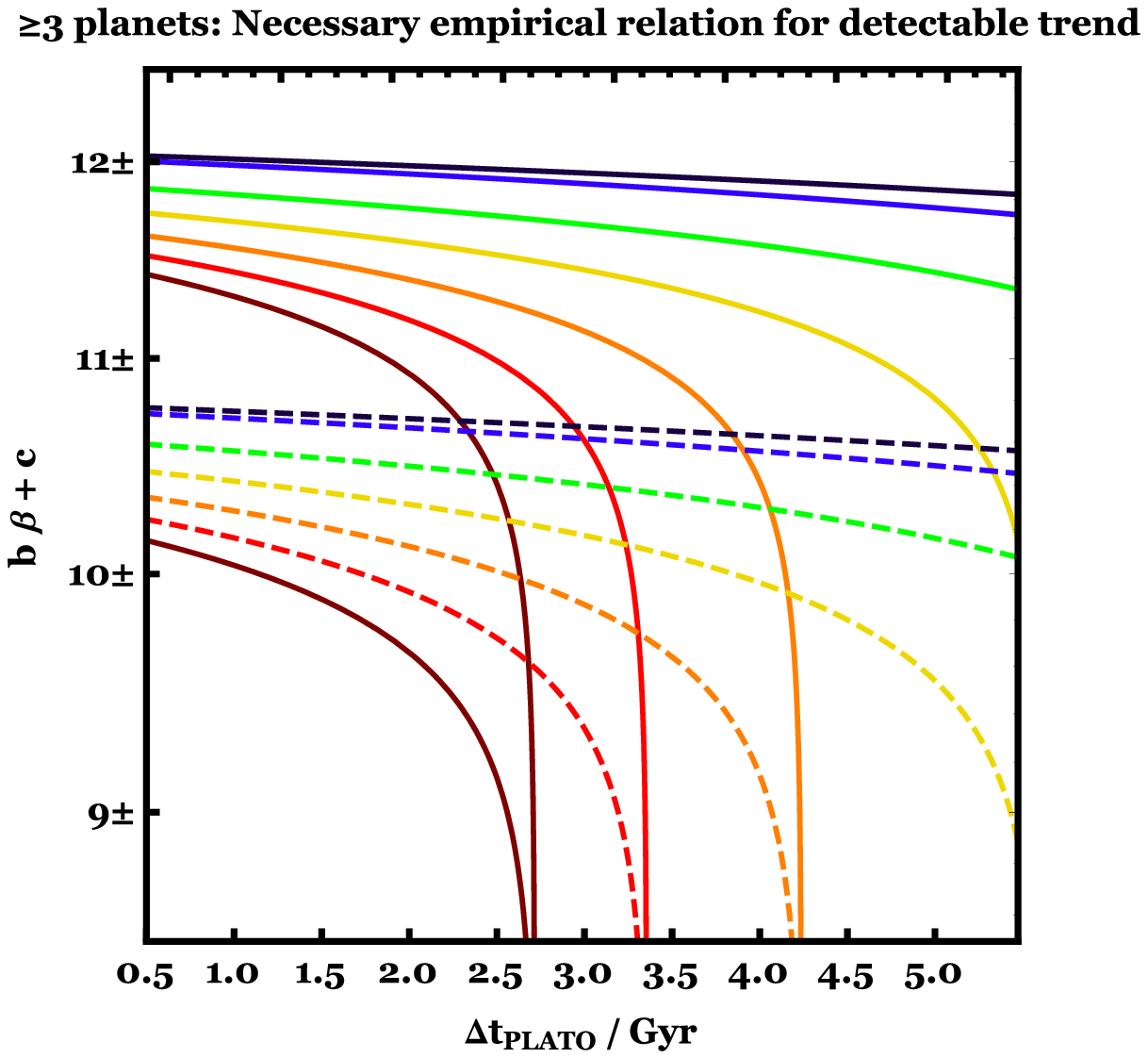,height=6.8cm,width=8.0cm} 
}
\centerline{
\psfig{figure=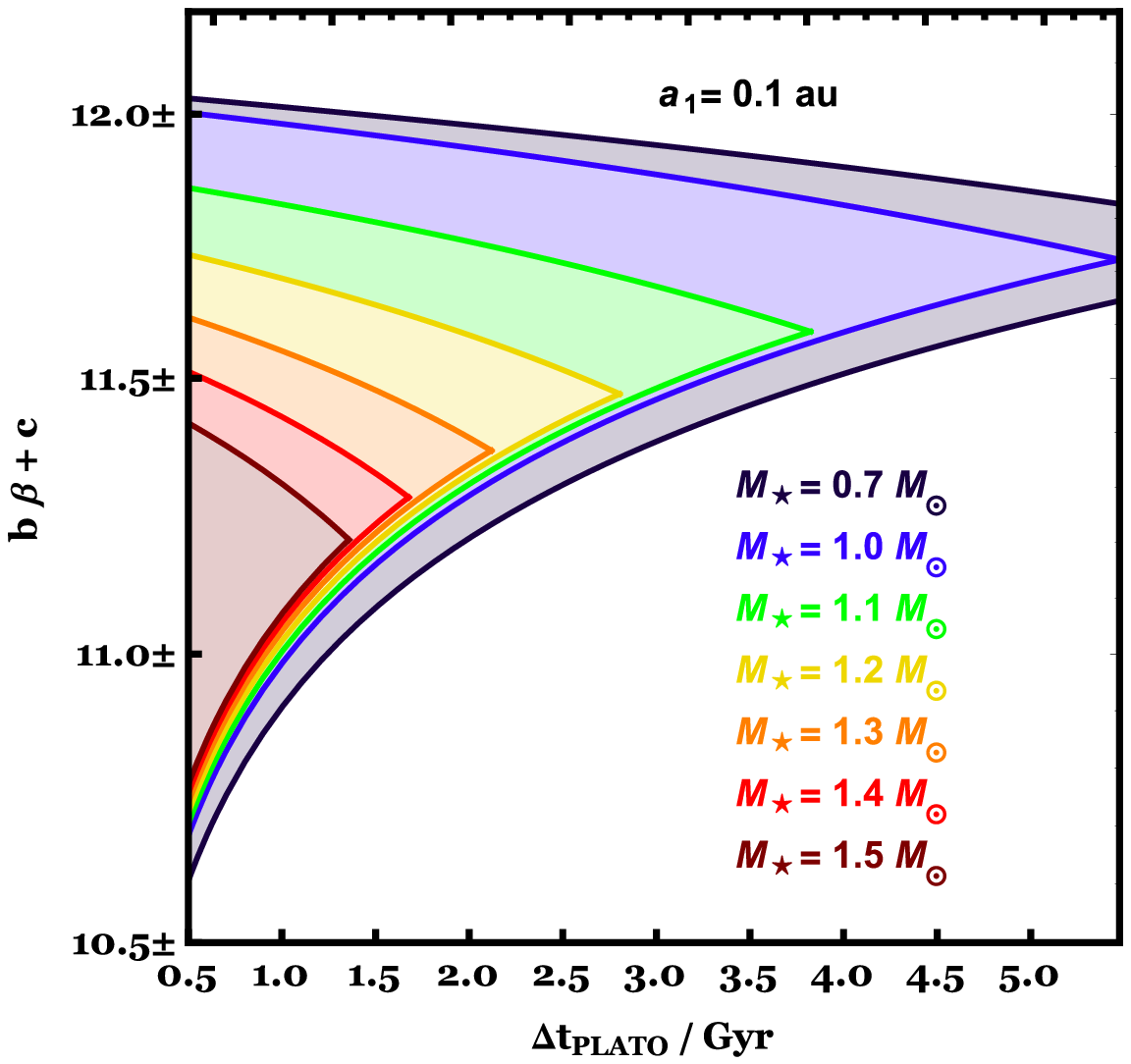,height=6.8cm,width=8.0cm} 
}
\caption{Values of the empirical planet instability relation 
$(b\beta + c)$ for systems 
with at least three planets for which {\it PLATO} will detect
a decreasing trend of planet frequency versus time.
The solid and dashed lines coloured lines in the upper panel
have the same designations as those in the upper panel of
Figure \ref{Plan2}. In the lower panel, the first time
bin is excluded, such that a detectable trend is possible only
in the shaded regions.  The $\pm$ symbols in the $y$-axis labels
indicate that the demarcated regions may be smeared out in
an unpredictable fashion.  If the packing parameter $\beta$ is too great,
then the systems will not produce a detectable trend.
}
\label{Plan3plus}
\end{figure}
%%%%%%%%%%%%%%%%% Figure eqmass

The upper panel of Figure \ref{Plan2} illustrates the minimum planet masses
for which {\it PLATO} will be able to detect a trend of decreasing
planet frequency with main sequence age,
provided that {\it PLATO} can discover enough planets at these
given masses to build up a large-enough sample.
From the {\it PLATO} primary sample of stars, we expect many 
thousands of terrestrial planets with 
orbital periods up to about one year to be detectable.  For a 
significant proportion of these planets, stellar 
activity signals will complicate the determination of their masses. However, by just 
crudely binning observations, {\it PLATO} will be able to recover many of these systems. 
Disentangling stellar activity signals is an ongoing area 
of research \citep[e.g.][]{basetal2014} that will, no doubt, further
improve the situation over the next decade.  

The first time bin is particularly important, and will contain a large sample of instabilities
with which to compare.  Consider the consequences of excluding that time bin; the lower panel 
of Figure \ref{Plan2} also imposes the restriction of Equation (\ref{min}).  If \textit{PLATO} were
to look for trends of planet frequency with stellar age only between the first and 
last time bins, then the planet mass range for which this technique would be successful is
less than an order of magnitude.

In any case, if we assume that {\it PLATO} will constrain stellar ages to within 10 per cent
for stars of spectral type K7 or earlier, then we can obtain some broad estimates
for the planet masses for which a decreasing frequency would be detectable.
Consider our model stars for which $M_{\star} \ge 1.0M_{\odot}$.
Their values of $t_{\rm MS}$ imply $\Delta t_{\rm PLATO} \lesssim 1$~Gyr
always.  Therefore, in the equal-mass case (upper panel), {\it PLATO} should be able
to detect a decreasing frequency trend for planet masses $\sim 10^{-2} M_{\rm J}$
and larger.  This statement is also true for $0.7M_{\odot} \le M_{\star} < 1.0M_{\odot}$
because of the flatness of the lowest-mass curve.  The bottom panel of 
Figure \ref{Plan2} also suggests that the trend is detectable
for planet masses as low as those found in super-Earths. {\it PLATO} will be able to 
detect these bodies, as well as even sub-Earth planetary radii, and thereby improve our
understanding of the currently poorly-constrained radius and mass 
distributions of the lowest mass planets\footnote{One exception is the 
current record holder for the lowest mass planet, PSR B1257+12A (sometimes known
as PSR 1257+12b), with a well-constrained mass of about $0.02M_{\oplus}$ \citep{wolszczan1994}, 
that resides in the first confirmed exoplanetary system \citep{wolfra1992}.}. 
Determining masses for the lightest planets is difficult. Recent studies, e.g. \cite{wulit2013} 
and \cite{weimar2014}, have shown potentially tantalising correlations
but remain somewhat controversial due to the lack of directly-measured planetary masses.

\subsection{Systems with three or more planets}

For more complex systems with higher numbers of planets, we cannot make
as direct a link with planetary mass as in the two-planet case.  Instead,
the key constraint is on the quantity $(b \beta + c)$.  By using
equation (\ref{max}) we obtain

\[
b \beta + c < 10.05 + \log_{10}{}
\bigg\lbrace
\left(
\frac{M_{\star}}{M_{\odot}}
\right)^{\frac{1}{2}}
\left(
\frac{a_1}{0.2 \ {\rm au}}
\right)^{-\frac{3}{2}}
\left(
\frac{\Delta t_{\rm PLATO}}{1 \ {\rm Gyr}}
\right)
\]

\begin{equation}
\ \ \ \ \ \ \ \ \ \ \ \ \ \ \ \ \ \ \ \ \ \ \ \
\times
\left(
\frac{t_{\rm MS}}{\Delta t_{\rm PLATO}} - 1
\right)
\bigg\rbrace
.
\label{bbetac}
\end{equation}

We emphasize that the applicability of Equation (\ref{bbetac})
is limited.  The intrinsic scatter in system stability
causes so-far-unpredictable deviations from this formula.
For example, Fig. 2 of \cite{smilis2009} indicates that
for a given set of $(b,c)$, unstable systems can occur
for different values of $\beta$ which vary by more than
unity.

We plot the curves resulting
from Equation (\ref{bbetac}) in the upper panel of 
Fig. \ref{Plan3plus}.  The $y$-axes labels 
deliberately contain the symbol $\pm$ in order to
indicate the uncertainty described in the last
paragraph. Equation (\ref{bbetac}) implies that 
if the system is too
widely separated initially (with $\beta$ too large), then
the instability will never occur quickly enough for a trend
to be detectable.  Alternatively, $\beta$ can never be too
small, because the resulting quick instabilities will be
included in the first time bin.  

If we exclude this time bin, then the result is the bottom
panel of Fig. \ref{Plan3plus}.  In this panel, the quantity
$(b \beta + c)$ is restricted to just about a few; too small of a
range to even encompass the different fits 
from multiple authors to instability timescales in a similar
region of parameter space \citep[see Fig. 3 of][]{puwu2015}.

Nevertheless, and despite large phase space of systems 
with three or more planets, we can make some concrete statements.  Consider
the study of \cite{smilis2009}, whose equation 6 is in the same form
as our Equation (\ref{comlog}) and who derive sets of $(b,c)$ based
on extensive suites of $N$-body numerical integrations.  They found
that for 5-planet systems where $\mu = 3.0035 \times 10^{-6}$ (corresponding
to Earth-mass planets orbiting a Solar-mass star), $(b,c) = (1.012, -1.686)$
holds for $3.4 \lesssim \beta \lesssim 8.4$.  These values correspond
to $(b \beta + c)\approx\left\lbrace 1.8,6.8\right\rbrace$, lower than
the ranges presented in Fig. \ref{Plan3plus}.  Consequently, instabilities produced
from that setup would occur within the first time bin.  

\cite{smilis2009}
also considered many other system types; for systems with three Earth-mass 
planets, they fit values of $(b,c) = (1.496, -3.142)$ for $\beta \ge 3.0$,
which would give $(b \beta + c)\approx \left\lbrace 10,11,12 \right\rbrace$ 
for $\beta \approx \left\lbrace 8.78,9.45,10.12 \right\rbrace$.  These
values of $\beta$ would place the systems in other time bins (besides the first).  
Comparing their 5-planet and 3-planet fits in their fig. 1 reveals that the functional
form of the fit might change for $\beta \gtrsim 8.5$.  Quantifying how, even
for just the equal-mass case, is difficult because of computational limitations.
Using $N$-body simulations to predict the very long term ($\sim$Gyrs) evolution of
transiting planets (such as those which will be discovered by \textit{PLATO}) 
pose an additional challenge because their semimajor axes would be on the scale 
of hundredths or tenths of an au 
(whereas \citealt*{smilis2009} adopted $a_1 = 1$ au).

\section{Summary}

Compared to previous experiments, the {\it PLATO} space mission will significantly 
improve robust stellar age estimates by developing separate internal mass distributions
for thousands of individual stars.  Consequently, {\it PLATO} will unveil   
currently undetectable trends in planetary systems.  In this work, we demonstrated
one such potential trend: decreasing planetary frequency with main sequence age for marginally
unstable equal-mass multi-planet systems.  For two-planet systems, this trend could be most secure
for planetary masses which are on the order of ten Earth masses, well within {\it PLATO}'s
planet detection capabilities.  Detection of this trend could also help constrain formation
mechanisms and test the theory \citep{puwu2015,volgla2015} that tightly packed inner planets
were more prevalent in the initial stages of the lifetime of planetary systems.

\section*{Acknowledgments}

We thank the referee for valuable comments on our manuscript.
DV benefited by support from the European Union through ERC grant number 320964.
AJM acknowledges support from grant number KAW 2012.0150 from the
Knut and Alice Wallenberg foundation and the Swedish Research Council (grant 2011-3991).

%\bsp

\label{lastpage}

\end{document}